\documentclass{pasa}%

\title[Upper limits on a radio halo in A3667 at 1.4 GHz]{Upper limits on a radio halo in Abell 3667 at 1.4 GHz}
\author[Johnston-Hollitt \& Pratley]{M. Johnston-Hollitt$^{1,2}$\thanks{Melanie.Johnston-Hollitt@vuw.ac.nz} \and L. Pratley$^1$\\
\affil{$^1$School of Chemical and Physical Science Victoria University of Wellington, PO Box 600, Wellington 6140, New Zealand}
\affil{$^2$Peripety Scientific Ltd., PO Box 11355 Manners Street, Wellington 6142, New Zealand}}%
\jid{PASA}
\doi{10.1017/pas.\the\year.xxx}
\jyear{\the\year}

\usepackage[authoryear]{natbib}
\bibpunct{(}{)}{;}{a}{}{,}
\usepackage{aas_macros}
\usepackage{hyperref} 
\usepackage{color}
\usepackage{graphicx}
\hypersetup{colorlinks,citecolor=blue,linkcolor=blue,urlcolor=blue}

\begin{document}%
\begin{abstract}
The presence of a radio halo in the massive, merging cluster Abell 3667 has recently become a focus of debate in the literature following a putative halo detection at 2.4 GHz despite a lack of detection at a range of lower frequencies between 120 MHz and 1.8 GHz. Here we develop a new method to place limits on radio haloes via generation of a realistic synthetic halo based on the brightness distribution of real haloes. The model generated extends on previous methods in the literature producing a single elliptical halo model, capable of being injected into mosaic as well as single observations. Applying this model to the deepest data available 1.4 GHz data for A3667 we derive an upper limit halo power of P$_{1.4} \leq 5.55 \times 10^{23}$ W Hz$^{-1}$. We discuss the result in the context of current scaling relation between the X-ray and radio properties of galaxy clusters and find that the lack of a halo in A3667 places the cluster on the border of the so-called `off-state' region in which clusters are expected not to host any diffuse emission. 
\end{abstract}
\maketitle%
\section{INTRODUCTION }
\label{sec:intro}

The large scale distribution of matter in the Universe can be described by a cosmic web of galaxies, consisting of galaxy groups, galaxy clusters, and filaments which have been observed to persist out to redshifts $\sim$2 \citep{Dehghan14}. In particular, the densest parts of the web are the intersections of cosmic filaments, and it is here that galaxy clusters, large groups of galaxies containing 100's to 1000's of galaxies embedded in a warm plasma known as the intra-cluster medium (ICM), form. The charged particles within the ICM emit via Bremsstrahlung radiation in the X-ray band and until recently X-ray surveys have been the mainstay for cluster detection \citep{Bohringer}. More recently instruments such as the Atacama Cosmology Telescope (ACT; \citealp{ACT}), the South Pole Telescope (SPT; \citealp{SPT}) and Planck \footnote{http://www.esa.int/Planck.} have paved the way for cluster detection via the Sunyaev-Zel'dovich (SZ) effect which allows the detection of mass limited cluster samples (e.g. \citealp{ACTcat,PSZ1,SPTcat, PSZ2}). In the radio there are a number of important phenomena associated with galaxy clusters which arise from synchrotron emission associated either with individual discreet objects in the cluster (star-forming galaxies or Active Galactic Nuclei, AGN) or from cluster-wide processes which give rise to diffuse emission in the form of relics and haloes. Relics are highly polarised arc-shaped emission regions on the edges of clusters thought to form when merger-generated shocks in the ICM accelerate charged particles within the cluster magnetic field via diffusive shock acceleration (DSA) \citep{Roettiger99}. The predicted spectral index profile for relics generated via DSA is a flattening on the outside edge followed by a steep spectral gradient pointed back towards the cluster centre and indeed this has been observed now in several systems, the first of which was the southern eastern relic in Abell 3667 (A3667) \citep{johnston-hollitt03} followed by the more widely known case of the Toothbrush relic \citep{vanWeeren10}, though several examples now exist \citep{Gasperin15,rvw17}. While the generation mechanism for relics is typically assumed to be DSA, recent examples have thrown that interpretation into doubt \citep{Vazza15,Botteon16,Akamatsu17}. Similarly in the case for radio haloes, which are roughly circular regions of diffuse emission co-located with the thermal X-ray emitting plasma at the heart of galaxy clusters, there are still open questions regarding the acceleration mechanisms at work. At present there are two possible mechanisms for the origin of radio haloes in clusters; the re-acceleration model also known as the primary model and the hadronic model also known as secondary model. In the re-acceleration model merger-driven turbulence re-accelerates primary cosmic ray electrons (CRes) in the magnetised ICM, generating synchrotron emission in the form of radio haloes \citep{2001ApJ...553L..15B, 2001MNRAS.320..365B, 2010ApJ...721L..82C}. Alternatively as relativistic cosmic ray protons (CRPs) will have effective proton--proton collisions in all galaxy clusters and this will produce ultra-relativistic electrons as a by-product of these interactions, the competing hadronic model \citep{1980ApJ...239L..93D, 1980PhDT.........1V, pfrommer08}, is also a possible mechanism for halo generation. The two models differ in both the strength of the generated haloes and the dynamical status of the clusters in which halo emission is predicted. 
The hadronic model predicts that all galaxy clusters will have faint radio haloes while the turbulent re-acceleration model predicts only those clusters undergoing a merging event will host a radio halo for some period after the merger and that the strength of the halo emission will be one to two orders of magnitude stronger. At present haloes are rare with only around 65 cases known \citep{Shakouri16} and these have seemingly all been detected in merging clusters at strengths comparable to those predicted in the turbulent re-acceleration model. Consequently, turbulent re-acceleration is currently favoured as the mechanism for halo generation in the literature \citep{BandJ2014}. However, the rarity of haloes in some of the most massive and recently merging clusters remains a puzzle. If the turbulent re-acceleration mechanism is correct we would expect to see haloes in all recently merging clusters but at present this is not the case and many clusters which exhibit key indications of recent merging across the electromagnetic spectrum do not have radio haloes. One such cluster is Abell 3667 which is known for its spectacular double relics \citep{johnston-hollitt03, hindson14, riseley15}, shocked X-ray gas \citep{Vikhlinin01,Finoguenov10} and spectroscopic confirmation of a plane of the sky merger \citep{mjh08,owers09}. Yet despite these indications of merging, there is not a clear halo detected for A3667. 
Here we present an attempt to search for a halo in A3667 at 1.4 GHz using deep Australia Telescope Compact Array (ATCA) data. While no such detection was made we place upper limits on possible halo power at 1.4 GHz via the injection of simulated radio haloes and provide an analytic model for elliptical halo profiles. Throughout we assume a $\Lambda$CDM cosmology with H$_{0}$=70 kms$^{-1}$Mpc$^{-1}$, $\Omega_{m}$=0.3 and $\Omega_{\Lambda}$=0.7. At the redshift of A3667, 0.0553 \citep{owers09}, 1 arc second is 1.074 kpc.

\section{A3667}
Over the last 50 years A3667 has been well studied over a large part of the radio spectrum ranging from 86 MHz to 5 GHz. Numerous interesting radio features have been detected, the most prominent of which are two large, bright radio relics located on the northwest (NW) and southeast (SE) periphery of the cluster. The NW relic has been known since the late 60s and imaged many times (e.g., \citealp{Mills61, Ekers69, Schilizzi75, Goss82, Rottgering97}) but these studies all failed to notice or properly consider the SE relic which is approximately 10 times fainter \citep{johnston-hollitt03,mjh04}. The first study of the entire cluster at high resolution (6"), including both the NW and SE relics, was carried out by \cite{johnston-hollitt03} at 1.4 GHz and 2.4 GHz with the ATCA using a combination of reprocessed archival data of the NW relic presented in \cite{Rottgering97} and new observations of the SE relic. In total \cite{johnston-hollitt03} used a combination of 30 pointings to create a mosaic of the full cluster spanning $\sim$ four square degrees. The data presented at both 6" and 43" resolution showed a greater extent to the previously known NW relic, characterised the properties of the SW relic and revealed a polarised filamentary structure in the relics similar to that recently seen in A2256 \citep{Owen14}. The average spectral index\footnote {Here we define the spectral index such that S $\propto\nu^{\alpha}$, where S is the flux density, $\nu$ is the frequency and $\alpha$ is the spectral index.} of the relics between 0.8 and 2.4 GHz was -0.9 $\pm$ 0.2 and -1.2 $\pm$ 0.2 for the NW and SE relics respectively, and as mentioned previously the spectral index of the SE relic exhibited the spatial gradient predicted by shock acceleration \citep{johnston-hollitt03}. More recently \cite{hindson14} investigated the relic emission at low radio frequencies (105 - 241 MHz) with the Murchison Widefield Array (MWA; \citealp{tingay13}) at a resolution of 3-5 arc minutes reporting a spectral index between 120 MHz and 1.4 GHz of -0.9$\pm$0.1 for both relics. Shortly after \cite{riseley15}  imaged A3667 with the KAT-7 telescope at 1.4 and 1.8 GHz with a resolution of 3 and 4 arc minutes, respectively. They report a consistent spectral index between 0.8 and 1.8 GHz for the NW relic of -0.8 $\pm$ 0.2 and a slightly flatter spectral index for the SE relic of -0.5 $\pm$ 0.2.

\subsection{Evidence for a Radio Halo in A3667}

The possibility of a radio halo in A3667 was first raised in \cite{johnston-hollitt03} who detected some patchy emission at the cluster centre in their deep 1.4 GHz image after a taper had been applied to produce a 43" resolution image. The total flux of the emission was reported to be 33 $\pm$ 6 mJy, however it was irregular and in three regularly spaced stripes in the positions where the side lobes of the two brightest objects in the field, MRC B2007-569 and J201258.890-570226.0, intersected. It was thus unclear if this emission was real or simply an imaging artefact. A decade later the detection of a putative halo and radio bridge connecting the NW relic to the centre of the cluster was reported by \cite{carretti13} using 2.3 and 3.3 GHz Parkes data after the subtraction of MRC B2007-569. Given the reported flux density at 2.3 GHz of the halo component of 44 $\pm$ 6 mJy with a peak brightness of 22 mJy beam$^{-1}$ and assuming a typical spectral index of -1.3, such a halo should have been readily detectable at low frequencies (see Section 4.3 of \citealp{hindson14} for a detailed discussion). However, MWA observations at 120, 149, 180 and 226 MHz found no support for the putative halo \citep{hindson14} and furthermore low resolution imaging at 1.4 GHz with KAT-7 has also failed to detect the halo \citep{riseley15}. 

\begin{figure}
\centering
\includegraphics[width=8.5cm]{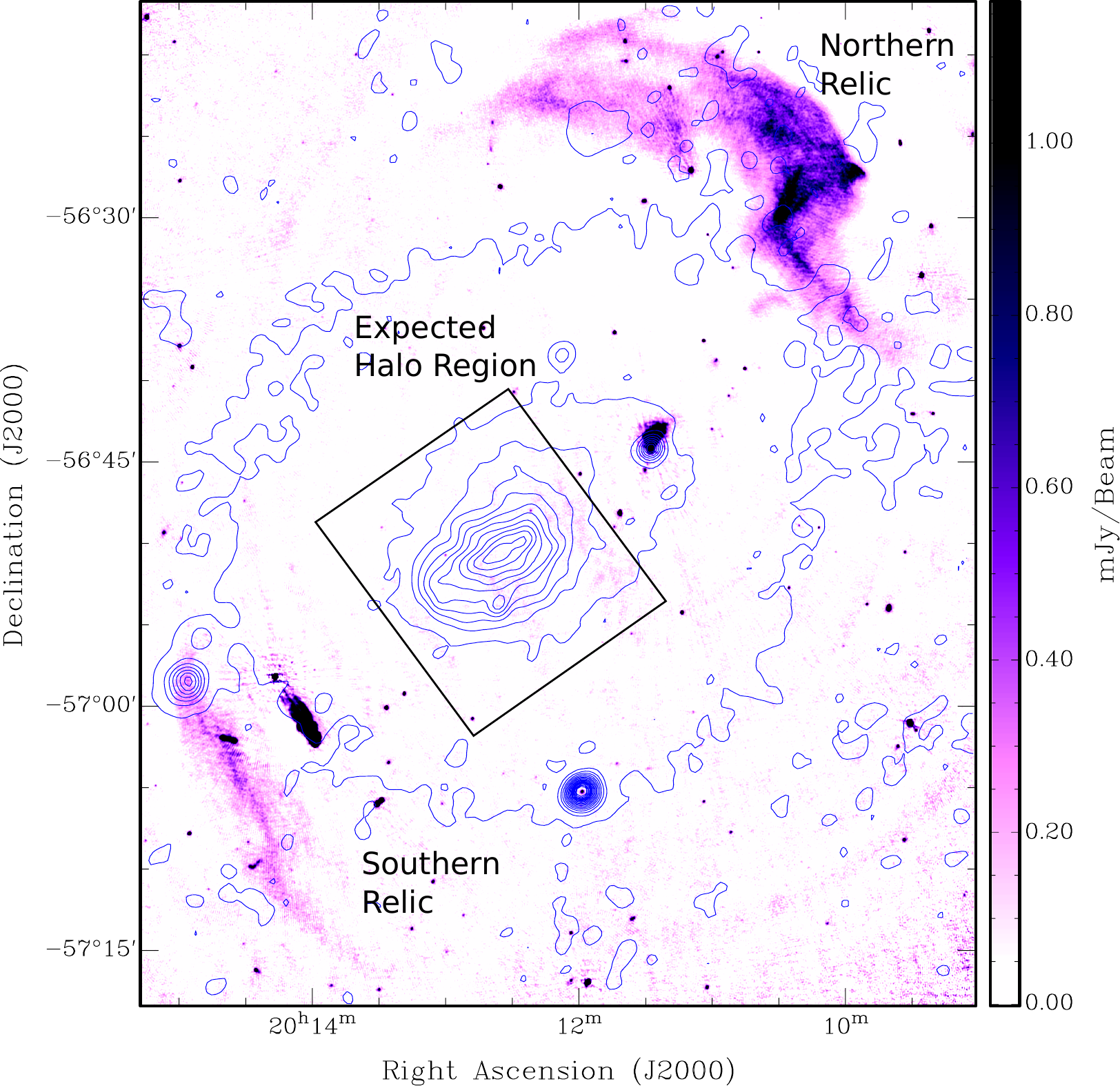}
\caption{30 pointing radio mosaic of A3667 at 1.4 GHz and full resolution (6") from \cite{johnston-hollitt03}. The blue contours show the ROSAT PSPC  X-ray contours.  Given that this is a merging cluster, if the theory of turbulent acceleration is correct, a radio halo is expected to be near where the X-ray is brightest, within the annotated box.}\label{fig:regionA3667}
\end{figure}

Given the now well established correlation between X-ray luminosity and the presence and power of radio haloes in merging clusters \citep{venturi07,brunetti09}, that A3667, which is one of the brightest known, merging X-ray clusters (L$_{x}$ = $8.74 \times 10^{44}$ erg/s \citealp{ebeling96}), is seen without a detectable halo in most radio images is a puzzle. Here we return to the deep 1.4 GHz dataset of \cite{johnston-hollitt03} to try to either uncover a radio halo in the expected location (see Figure \ref{fig:regionA3667}) or place robust upper-limits on the presence of a halo.

\section{1.4 GHz Imaging}
As mentioned above, \cite{johnston-hollitt03} presented a 30-pointing mosaic of A3667 comprised of reprocessed archival and new ATCA data collected 1.4 and 2.4 GHz. The observations were collected over a 9 year period from 1993 to 2002 using eight different array configurations to better fill the uv-plane. They represent the most comprehensive and detailed imaging of A3667 yet undertaken. Full details of the observation and calibration used here are described in detail in \cite{johnston-hollitt03}. In this work we used the fully reduced visibility data for each pointing, with no additional reduction of the data. 

\subsection{Removal of Central Point Sources}

To improve on the possible halo detection reported in \cite{johnston-hollitt03}, we first removed all of the detectable point sources in the central part of the cluster via modelling and subtracting them from the visibility data. As each source will appear at a different position from the phase centre of each pointing and thus have different attenuation due to the antenna response, removal of sources had to be performed individually for each of the pointings. In total 45 point sources were removed or partially removed from all 30 pointings via the task UVMODEL in MIRIAD \citep{sault95}. In addition we attempted to model and partially remove both bright tailed radio galaxies so as to reduce the side lobe pattern in the centre of the field. The data were then mosaiced together via a joint deconvolution to form a final, tapered image at a resolution of 43". While this image considerably reduced the intersecting grating lobes at the centre of the cluster, the root-mean-squared noise of 1.3 mJy beam$^{-1}$ was higher than in the original image by \cite{johnston-hollitt03}, nevertheless there was no evidence of a halo present confirming the suspicion that the putative halo in these data reported by \cite{johnston-hollitt03} was an imaging artefact. 
 
\subsection{Determining the Halo Upper Limit}
\label{sec:dist}
Having been unable to establish the presence of a halo via direct imaging, we now use the visibility data after removal of central bright sources to set a robust upper limit on the 1.4 GHz power of a halo in A3667.

A method similar to \cite{brunetti07} was used to model the synthetic halo, where the integrated radial flux profiles of real haloes were used to determine the brightness distribution of the synthetic halo, which is then scaled to the appropriate physical size for the cluster under investigation \citep{venturi07,venturi08}. However, whereas previous processes have modelled the halo profile as a series of circular Gaussians of up to five components of different sizes derived to approximate the observed average halo brightness profile (Venturi, private communication), here we introduce a new technique to provide a single continuous model. We provide our fit to the cumulative halo flux as a function of radius from the halo centre for others to use. Furthermore, we  extend the method from circular to elliptical haloes, allowing better matching of the halo parameters with the X-ray morphology seen in clusters. 

\begin{figure}
\centering
\includegraphics[width=9cm]{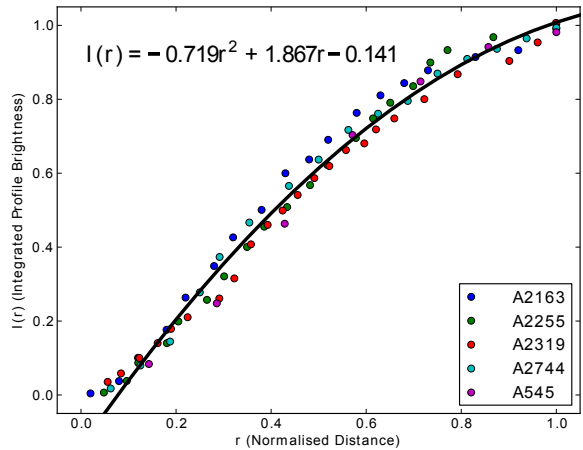}
\caption{The normalised integrated radial flux profiles of five radio haloes from the literature (A2319, A545, A2744, A2163, A2255) and the resultant average polynomial fit (average of the coefficients). The derivative of this fit, $\frac{dI}{dr}$, was used to model the radial brightness of the synthetic halo used here (see Equation \ref{eq:dist}).}
\label{fig:profile}
\end{figure}

As in \cite{brunetti07} we used the integrated radial flux profiles for the haloes in several galaxy clusters (A2319, A545, A2744, A2163, and A2255), scaled to a common size and intensity to derive our halo brightness profile. A linear least squares fit was performed on the normalised profiles to obtain a realistic scalable radial brightness distribution (see Figure \ref{fig:profile}). We found the best fit for the average halo profile to be given by the polynomial $I(r)= -0.719r^2 +1.867r -0.141$. Using this polynomial fit, the brightness within a pixel was assumed to be 
\begin{equation}
B(r)=(\mbox{Total Integrated Flux})\times \left( \frac{d I}{d r} \right)\times \frac{\Delta A}{2\pi}
\label{eq:dist}
\end{equation}
where $r$ is the normalised distance of the pixel from the centre of the circle and $\Delta A = \Delta r \Delta \theta$ is the area of the pixel. 

To extend this result to elliptical haloes we simply change the coordinates of $r$ in the $xy$-plane, to:

\begin{equation}
\resizebox{.5 \textwidth}{!} 
{
    $ r(x,y)=\sqrt{\left( \frac{( x_{0}-x) \cos\beta +( y_{0}-y) \sin\beta}{A} \right)^2 +\left(\frac{(y_{0}-y)\cos\beta -(x_{0}-x)\sin\beta}{B}\right)^2}
 $
}
\label{eq:ellipse}
\end{equation}

Here $A$ and $B$ are the semi-major and semi-minor axes of the ellipse, $\beta$ is the angle of rotation and $(x_0,y_0)$ is the coordinate centre for the ellipse. This change of coordinates can be thought of as scaling the normalised $r$ value depending on the orientation and dimensions of the ellipse, similar to a substitution in an integral. This means the total flux of the synthetic halo is conserved (of course $\Delta r$ is now different in these new coordinates, but it can be simply calculated for each pixel).

It is important to note that interferometers do not sample the entire Fourier plane and this has consequences on the amount of flux recovered from the final image as compared to the original model. In particular, when the halo is degridded in the $uv$-plane, some Fourier components, corresponding to those regions of the plane not sampled by the interferometer are removed, meaning the total flux is not conserved as we loose power from the model when omitting the non-sampled components of the power spectrum. Furthermore, the deconvolution process used in standard radio interferometric imaging `CLEAN' does not fully recover flux of low signal-to-noise sources as they are often not included in the set of deconvolved sources. As a result when injecting synthetic haloes at low surface brightness into $uv$-data and then re-imaging the data, the detected flux will be an upper limit governed by our choice of $uv$-coverage and the selection of deconvolution algorithms. To get around this practical problem, the profile of the model halo was normalised to have a total integrated flux of our choosing. Then the model of the halo was degridded and the resultant estimated halo visibilities added to the measurements. 

\subsection{Application to Mosaics}
\label{sec:scale}
All previous attempt to provide upper limits for radio halo detections have dealt with injecting a synthetic halo into a single pointing  \citep{brunetti07, venturi07, venturi08, kale13, kale15}. However, here we must undertake the process on a mosaic meaning, as with the removal of contaminating sources in the visibility data, we need to inject the synthesis halo on a pointing by pointing basis, properly accounting for the change in brightness distribution for each pointing due to the attenuation of the telescope across the field of view (primary beam). Thus, when injecting the synthetic halo into the visibilities for each image, the intensity should scale inversely with the distance away from the phase centre of the image. Here this attenuation has been modelled using the analytic expression for the ATCA's response as a function of radius, given in Equation \ref{eq:response}, with values of $a_1=8.99\times10^{-4}$, $a_2=2.15\times10^{-6}$, $a_3=-2.23\times10^{-9}$, $a_4=5.56\times10^{-12}$ \citep{wieringa92}. 

Obviously, the ATCA's attenuation pattern is similar but not identical to the response of other interferometers and thus one should use the appropriate curve when considering this approach on mosaic data taken with different instruments. 

\begin{equation}
A(r)=\frac{1}{1+a_1 r^2+a_2 r^4 +a_3 r^6 +a_4 r^8}
\label{eq:response}
\end{equation}

\subsection{Synthetic Halo for A3667}
Having established in Sections \ref{sec:dist} and \ref{sec:scale} the general brightness distribution and scaling required as a result of using different pointings of a mosaic, we proceed to inject the synthetic halo signal into the A3667 1.4 GHz data. In order to follow a realistic model of the halo, we choose to model the halo as an ellipse using Equation \ref{eq:ellipse}, with $A=420$  kpc, $B=300$ kpc, and $\beta = \frac{\pi}{4}$. 
Again using the task UVMODEL in MIRIAD we calculated individual models for each pointing of the mosaic so as to place the halo in the X-ray brightest part of the cluster accounting for the correct attenuation and brightness scale. 

\begin{figure}
\centering
\vspace{-0.9cm}
\includegraphics[width=8cm, angle=-90]{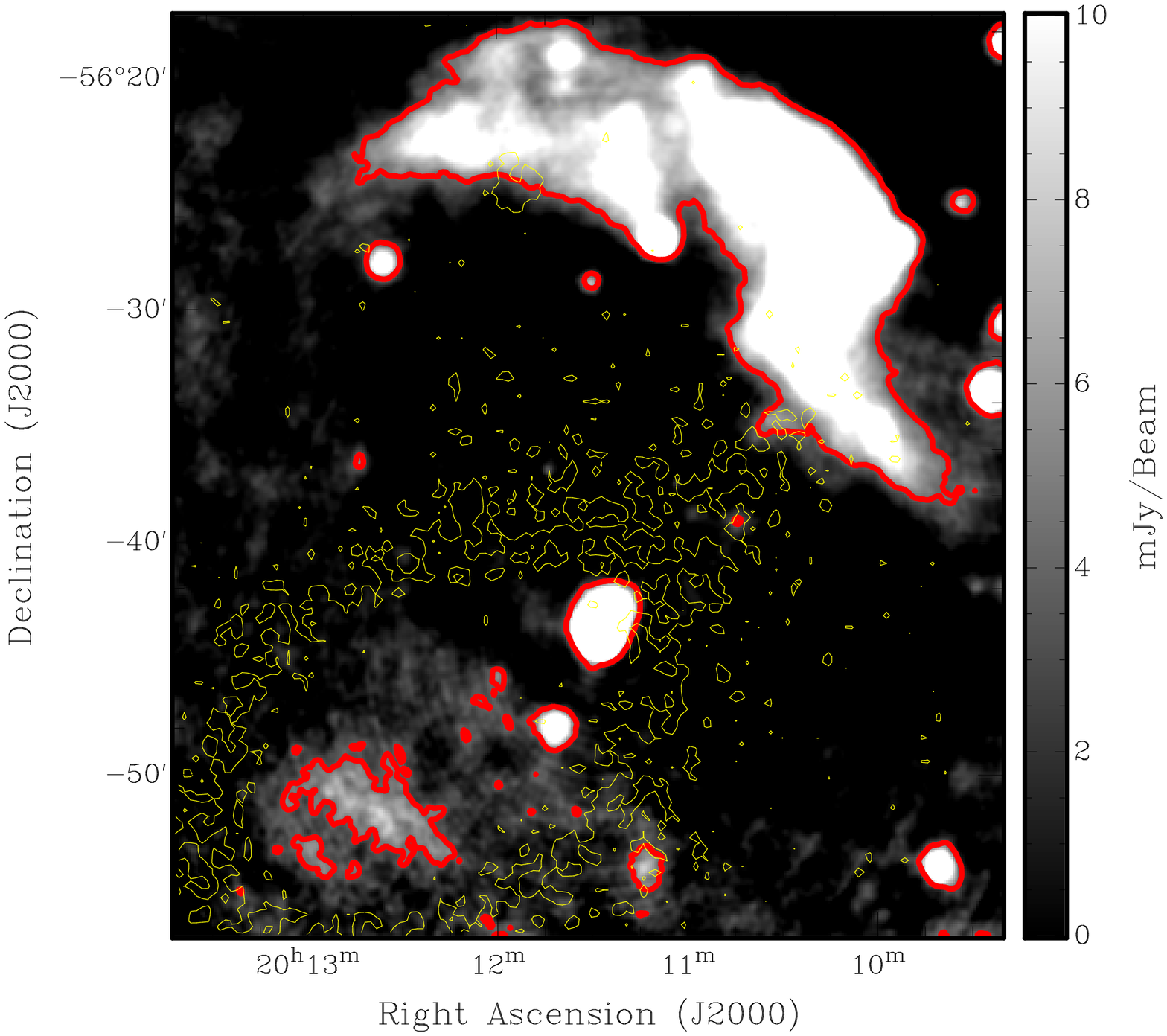}
\includegraphics[width=8cm, angle=-90]{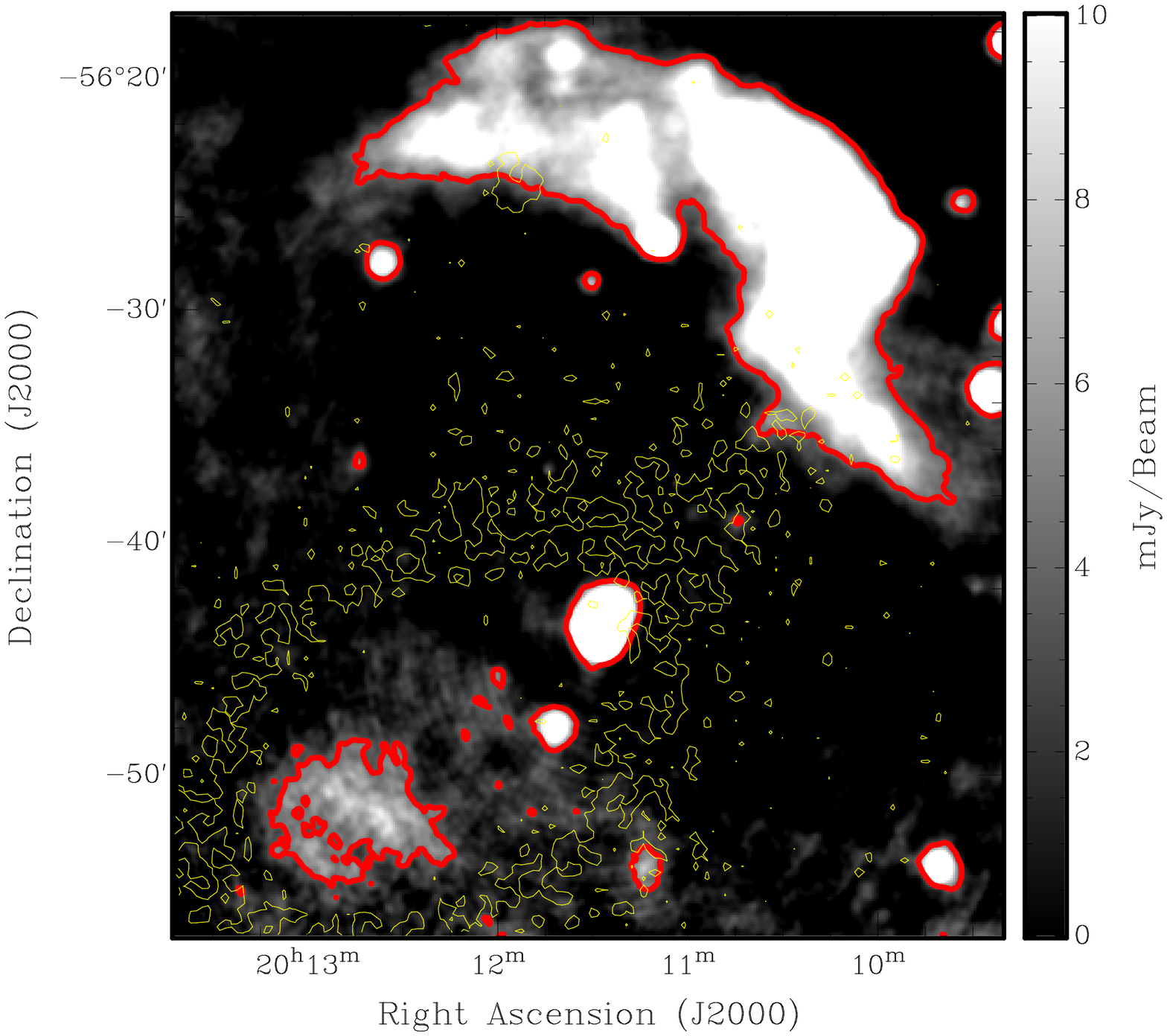}
\includegraphics[width=8cm, angle=-90]{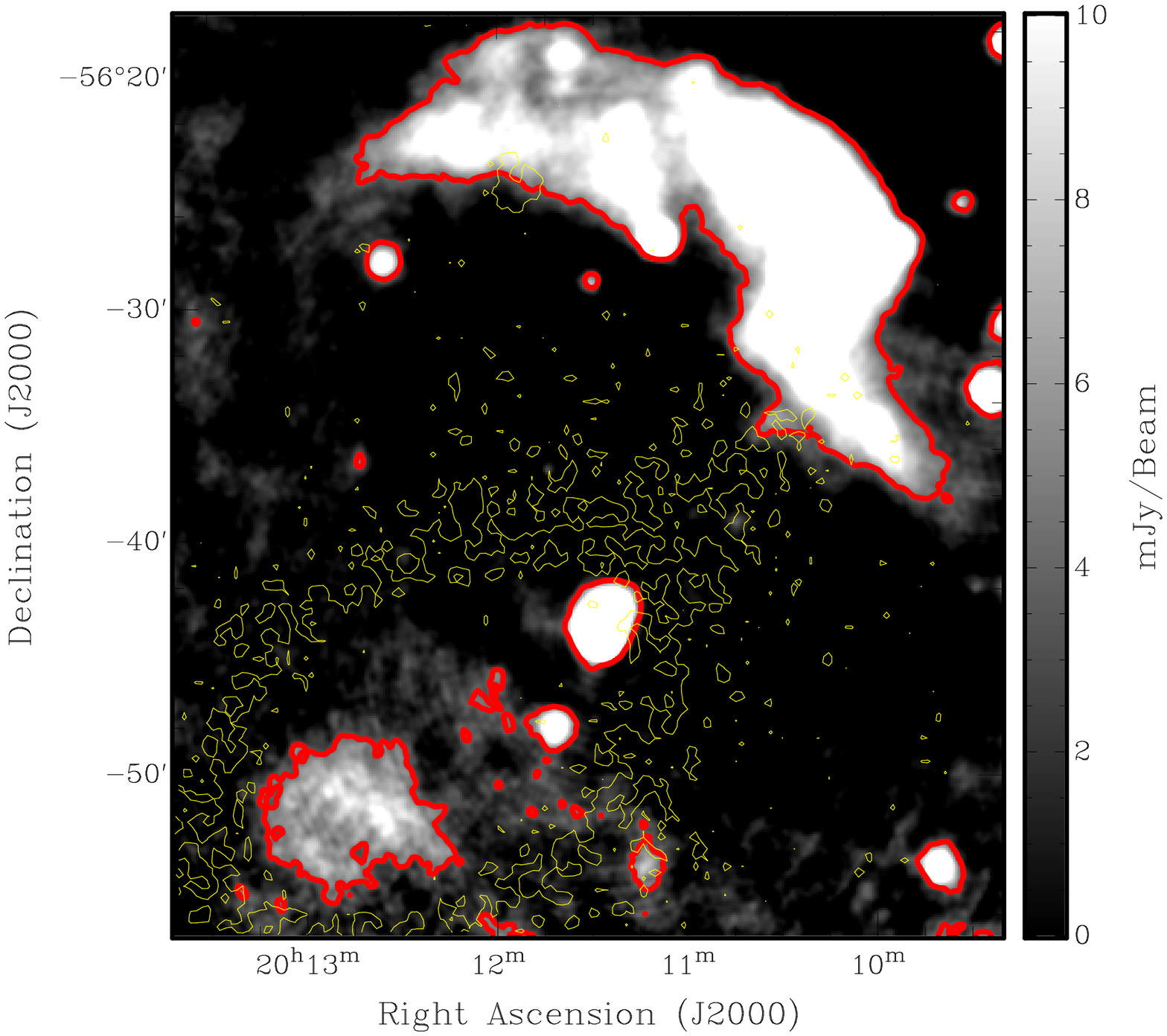}
\caption{Results of injecting a synthetic halo with different total flux densities into the 1.4 GHz A3667 mosaic tapered to 43" resolution. The halo is placed at the centre of the X-ray emission with approximately the same elliptical shape. The three sigma detection threshold for this image (4 mJy beam$^{-1}$) is shown with red contours while the yellow contours represent the X-ray emission. Each image has a different initial total integrated flux density for the synthetic halo being: 50 mJy, 75 mJy \& 100 mJy from top to bottom, respectively.}
\label{fig:fakehalores}
\end{figure}

We injected haloes with three different total flux densities: 100, 75, and 50 mJy and re-imaged the entire A3667 mosaic at the same resolution of 43". 
Figure \ref{fig:fakehalores} gives the results after application to the mosaic.

One of the difficulties with injecting haloes to determine an upper limit is knowing when to declare a detection and how to make this objective rather than subjective. There seems to be no standard defining when an injected halo is detected in the literature with various authors applying either undefined or subjective (by eye) methods. Here we elect to define a detection when at least 25\% of the surface area of the initial halo was recovered above 3 times the root mean squared (rms) noise in the image ($3\sigma$). The standard deviation, $\sigma$, of the near by noise was 1.3 mJy beam$^{-1}$, making the detection limit 4 mJy beam$^{-1}$. In the three cases we considered we recovered 16.1\%, 25.1\% and 33\% of the surface area above 3$\sigma$ for haloes with a total flux of 50 mJy, 75 mJy, and 100 mJy, respectively. Thus by our definition 75 mJy is the upper limit on the total 1.4 GHz flux density for the detection of a halo in A3667. Assuming a spectral index of -1.3, this corresponds to a k-corrected limit on the power of the halo of P$_{1.4} \leq 5.55 \times 10^{23}$ W Hz$^{-2}$. 

 \cite{carretti13} extrapolated from their 3.3 GHz Parkes data assuming a spectral index of -1 to obtained a 1.4 GHz power for their putative halo of P$_{1.4} = 7.5 \times 10^{23}$ W Hz$^{-2}$. Adjusting this to a spectral index of -1.3 as used here, this would give a limit of P$_{1.4} = 8 \times 10^{23}$ W Hz$^{-2}$. These values are both higher than the upper limit presented here, making it hard to reconcile their putative halo detection with this work. In fact the comparison between their putative detection and the limit set here suggest the spectral index of the emission would have to be considerably flatter than is typical for a radio halo. This might suggest the emission seen in the Parkes data is Galactic in origin which would be consistent with the MWA data that demonstrates A3667 sits on a ridge of diffuse Galactic emission \citep{hindson14}.

\section{Discussion}

\begin{figure}
\vspace{-0.4cm}
\centering
\includegraphics[width=8.5cm]{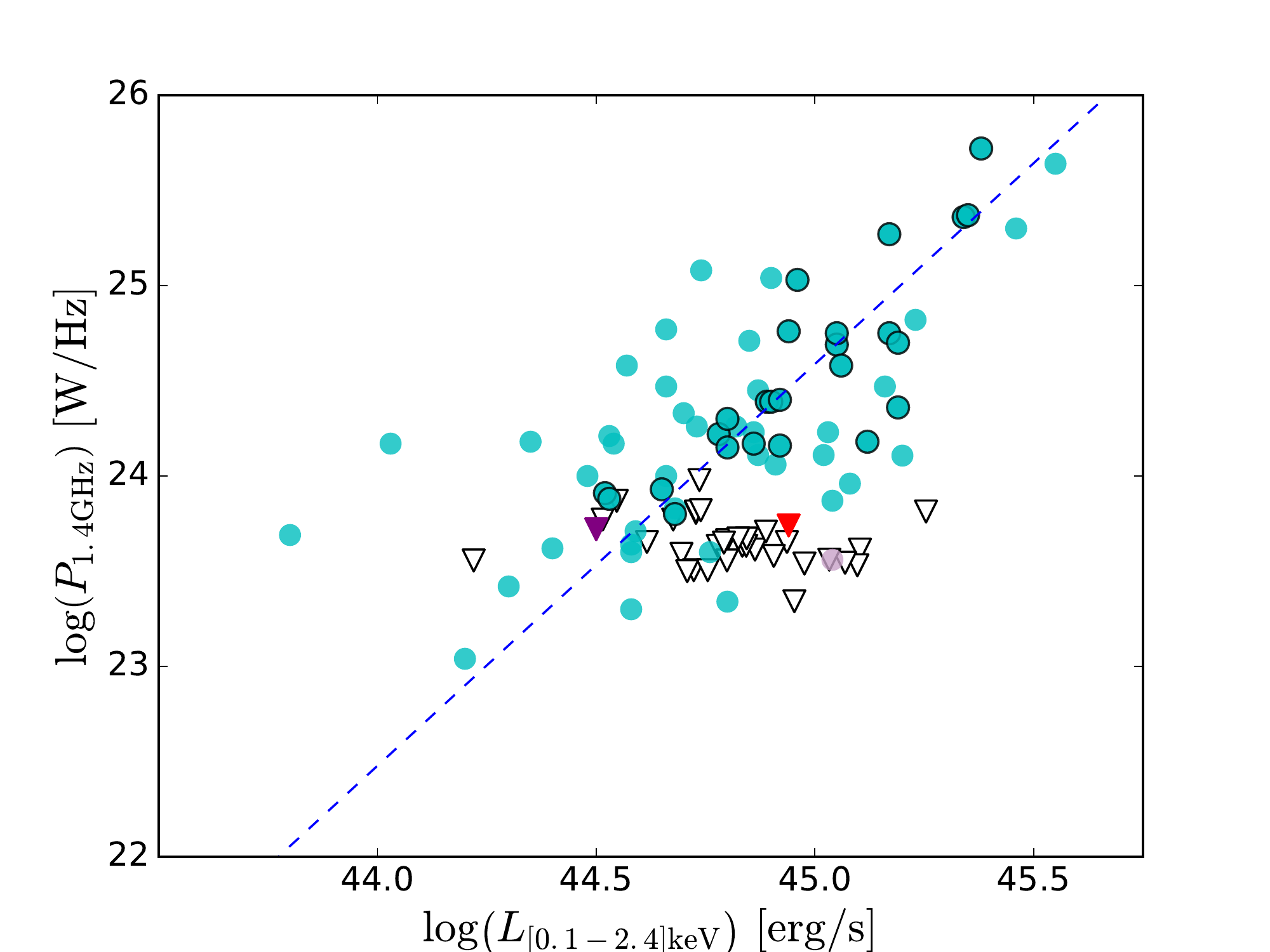}
\vspace{-0.3cm}
\includegraphics[width=8.5cm]{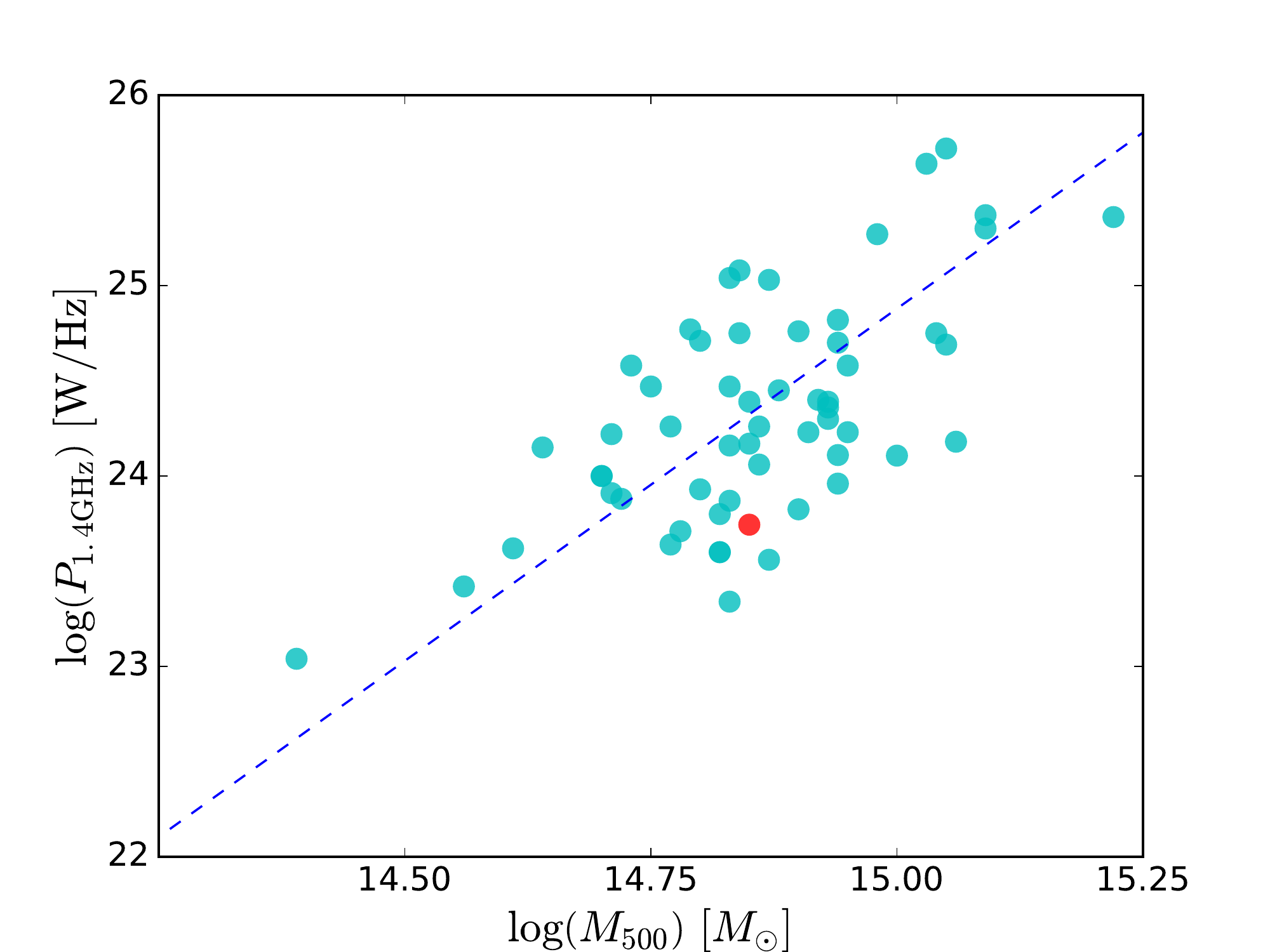}
\label{fig:venturi}
\vspace{-0.5cm}
\caption{Top: The $\log P_{1.4 GHz}$ vs. $\log L_{[0.1-2.4]}$ plot showing the correlation between the total halo power and X-ray power of clusters. The aqua dots are the 66 known haloes in the literature while those with black outlines are the 25 used in the correlation fit of \cite{cassano13} which is shown by the dashed line. The red filled triangle represents the upper limit power for a halo in A3667, the unfilled triangles represent upper limit halo powers for clusters from the literature (black values are from the extended GMRT Radio Halo Survey \citep{venturi07, venturi08, kale13, kale15} and the purple value is from \citep{russell11}).  The lilac dot is the only halo \citep{bonafede15} found in the `so-called' off-state region defined by  \cite{2011ApJ...740L..28B}. Despite that A3367 is a well-known merging cluster, it sits considerably below the \cite{cassano13} correlation, right on the border of the off-state region. Bottom: The $\log P_{1.4 GHz}$ vs. $\log L_{[M_{500}]}$ plot showing the correlation between the total halo power and M$_{500}$ mass. The aqua dots are the 57 known haloes in the literature with SZ-derived mass, the correlation fit of \cite{cassano13}, again using a subset of 25 clusters, is shown by the dashed line, and A3667 is marked by a red dot. }
\end{figure}

It has long been known there is a relationship between the X-ray properties of galaxy clusters and the power of radio haloes detected in some clusters. This was first expressed as a correlation between the cluster X-ray-derived temperature of the cluster and the 1.4 GHz radio halo power \citep{liang}. More recently this relationship has been expressed as a correlation between the X-ray luminosity and 1.4 GHz radio power \citep{cassano13}.  The top panel of Figure \ref{fig:venturi} shows the upper limit for a halo in A3667 on the P$_{1.4 GHz}$ vs.  L$_{x}$ diagram showing all 66 known haloes in the literature \footnote{Note that we do not include Abell 1213 as we do not consider this a halo. See Duchesne et al. (in prep) for further discussion}. We find the upper limit halo power for A3667 is located with other upper limits for non-detections, far off the correlation for real haloes found in merging clusters. In fact A3667 lies on the border of the so-called `off-state' region in which clusters are assumed to be radio quiet hosting neither haloes or relics \citep{2011ApJ...740L..28B}.  The bottom panel of Figure \ref{fig:venturi} gives the position of A3667 in the comparison between halo power and cluster mass for the subset of 57 clusters hosting radio haloes for which SZ-derived masses are currently available. Here it seems A3667 is in a regime consistent with the masses of systems in which we expect a reasonable fraction ($\sim$30 per cent) of clusters to host a halo \citep{cassano12}.

Of the 66 haloes known to date in the literature (dots on Fig. \ref{fig:venturi}), all of them have been associated with merging systems. This association with mergers and the lack of detectable gamma-ray emission in clusters generally, strongly favours the scenario that radio haloes are generated in massive clusters via turbulent re-acceleration. Conversely the clusters for which no halo has been detected are more relaxed, though some signatures of merging can be detected, and in some cases strong merging signatures are found despite the lack of presence of diffuse emission (e.g. A2146; \citealp{russell11}). Thus it seems merging is a necessary but not sufficient condition for halo generation and the difference between merging systems with and without diffuse emission may well be the presence or otherwise of a suitable electron population to accelerate \citep{ensslin01,mjh17}. While it is easy then to understand how clusters without a suitable electron population can undergo a merger and not produce diffuse radio emission as in A2146 \citep{russell11}, it is more of a puzzle as to how a cluster like A3667 which is undergoing a merger and clearly has electrons to reaccelerate, as demonstrated via the presence of radio relics, is without a halo. Even considering that simulations suggest halo powers will vary over the lifetime of a merger \citep{donnert13}, eventually ceasing to radiate at 1.4 GHz, that the relics in A3667 are still visible suggests that this point has not yet been reached for this cluster. However, given the considerable scatter in the P$_{1.4 GHz}$ vs.  L$_{x}$ plane, it is possible that a weak halo sits just below the current detection threshold. Some evidence for this argument comes from the MWA observations processed to favour large-scale emission \citep{hindson14} which show that the cluster sits atop a large ridge of Galactic emission which may be obscuring a faint halo, thus future observations with the upgraded MWA or eventually SKA LOW may yet detect a halo.

\section{Conclusion}
We present the most detailed attempt to detect a radio halo in A3667 at 1.4 GHz to date. Having removed 45 point sources and partially removed the two brightest extended sources in the field, we re-imaged the cluster using the data presented in \cite{johnston-hollitt03}. We find no evidence for a halo in these data and conclude that the putative halo at 1.4 GHz previously reported by  \cite{johnston-hollitt03, mjh04} was an imaging artefact. 

In order to determine an upper limit to a possible halo we develop a new halo generation technique in which the halo is modelled as a continuous distribution following a realistic brightness profile which can be configured as any elliptical shape. This new technique is further expanded to allow the use of this method on mosaic images via simply accounting for the attenuation response of the instrument as a function of pointing. Using this method, we place an upper limit on the 1.4 GHz power for a radio halo in A3667 as P$_{1.4} \leq 5.55 \times 10^{23}$ W Hz$^{-2}$, which places the cluster on the edge of the so-called `off-state' region of the power versus X-ray luminosity plane. Given that A3667 is a known merging cluster, with known radio relics and thus a ready populations of electrons to re-accelerate, the lack of a halo is problematic for the turbulence re-acceleration model. However, we note the large scatter in the P$_{1.4 GHz}$ vs.  L$_{x}$ plane might suggest that a weak halo is still present, just below our current detection limits. Future objections with the MWA II or SKA LOW should resolve this issue.  

Finally, we note that while this paper was under review a similar, but not identical, approach to more realistic halo generation became publicly available \citep{bonafede17}. In this work the authors model the halo brightness profile as an exponential function with some brightness fluctuations, rather than the measured brightness profile used here. Using an exponential distribution also allows for elliptical halos in a straightforward manner. That these two works appeared independently over the same time period highlights the need of the community for more realistic synthetic halo models. As we move towards widefield surveys on instruments such as the Australian SKA Pathfinder (ASKAP) or the MWA, the need to account for injection processes in mosaiced images, as presented here, will increase.

\begin{acknowledgements}
MJ-H acknowledges support from the Marsden Fund which is administered by the Royal Society of New Zealand and the Institute of Advanced Studies, University of Bologna for an ISA Senior Fellowship which allowed the opportunity to complete this long over-due manuscript. LP was supported in this work by a Victoria Graduate Scholarship. The Australia Telescope Compact Array is part of the Australia Telescope National Facility which is funded by the Australian Government for operation as a National Facility managed by CSIRO. This research has made use of the NASA/IPAC Extragalactic Database (NED), which is operated by the Jet Propulsion Laboratory, California Institute of Technology, under contract with the National Aeronautics and Space Administration.
\end{acknowledgements}

% UNCOMMENT THE LINES BELOW IF YOU WISH TO USE BIBTEX
\bibliographystyle{apj}
\bibliography{pasa_refs}

\end{document}